\documentclass{aa}
\usepackage{pifont}
\usepackage{amsmath}
\usepackage{graphicx}
\usepackage{txfonts}
\begin{document}
   \title{Asteroseismic modelling of the metal-poor star $\tau$ Ceti}

   \author{Y. K. Tang
             \inst{1,2}
             and
           N. Gai
          \inst{3,4}}
   \institute{
  Department of Physics, Dezhou University, Dezhou 253023, P. R. China\\e-mail: tyk450@163.com
  \and
  Key Lab of Biophysics in Universities of Shandong, Dezhou 253023, P. R. China\\
  \and
  Department of Astronomy, Beijing Normal University, Beijing 100875, P. R. China\\ e-mail: gaining@mail.bnu.edu.cn
 \and
 Department of Astronomy, Yale University, P.O. Box 208101, New Haven, CT 06520-8101, USA }

  \abstract
   {Asteroseismology is an efficient tool not only for testing stellar
structure and evolutionary theory but also constraining the
parameters of stars for which solar-like oscillations are detected,
presently. As an important southern asteroseismic target, $\tau$
Ceti, is a metal-poor star. The main features of the oscillations
and some frequencies of $\tau$ Ceti have been
 identified. Many scientists propose to
 comprehensively observe this star as part of the Stellar Observations Network Group. }
   {  Our goal is to obtain the optimal model and reliable fundamental parameters for the metal-poor star $\tau$
       Ceti by combining all non-asteroseismic observations with these seismological data.}
   { Using the Yale stellar evolution code (YREC), a grid of stellar model candidates
    that fall within all the error boxes in the HR diagram have been constructed, and both
    the model frequencies and large- and  small- frequency separations
  are calculated using the Guenther's stellar pulsation code.  The $\chi^{2}_{\nu c}$ minimization is
  performed to identify the optimal modelling
    parameters that reproduce the observations within their errors.
    The frequency corrections of near-surface effects to the calculated frequencies using the empirical law,
     as proposed by Kjeldsen and coworkers, are applied to the models.}
   {We derive optimal models, corresponding to masses of about 0.775 -- 0.785 $M_{\odot}$ and ages of about 8 -- 10
Gyr. Furthermore, we find that the quantities derived from the
non-asteroseismic observations (effective temperature and
luminosity) acquired spectroscopically are more accurate than those
inferred from interferometry for $\tau$ Ceti, because our optimal
models are in the error boxes B and C, which are derived from
spectroscopy results.}
  {}

\titlerunning{Asteroseismology model of the metal-poor star $\tau$ Ceti}
\authorrunning{Y. K. Tang et al.}

   \maketitle

%

\section{Introduction}
  The solar five-minute oscillations have led to a wealth of information
about the internal structure of the Sun. These results have
stimulated various attempts to detect solar-like oscillations for a
handful of solar-type stars. Solar-like oscillations have been
confirmed for several main-sequence, subgiant and red giant stars by
the ground-based observations or by the CoRoT and the Kepler space
missions, such as  $\nu$ Indi (Bedding et al. 2006; Carrier et al.
2007), $\alpha$ Cen A (Bouchy \& Carrier 2002; Bedding et al. 2004),
$\alpha$ Cen B (Carrier \& Bourban 2003a; Kjeldsen et al. 2005),
$\mu$ Arae (Bouchy et al. 2005), HD 49933 (Mosser et al. 2005),
$\beta$ Vir (Marti\'{c} et al. 2004a; Carrier et al. 2005a), Procyon
A (Marti\'{c} et al. 2004b; Eggenberger et al. 2004a;  Arentoft et
al. 2008; Bedding et al. 2010), $\eta$ Bootis (Kjeldsen et al. 2003;
Carrier et al. 2005b), $\beta$ Hyi (Bedding et al. 2001, 2007;
Carrier et al. 2001), $\delta$ Eri (Carrier et al. 2003b), 70
Ophiuchi A (Carrier \& Eggenberger 2006), $\epsilon$ Oph (Ridder et
al. 2006), CoRoT target HR7349 (Carrier et al. 2010), KIC 6603624,
KIC 3656476 and KIC 11026764 (Chaplin et al. 2010), etc.
Furthermore, the large and small frequency separations of p-modes
can provide a good estimate of the mean density and age of the stars
(Ulrich 1986, 1988). On the basis of these asteroseismic data,
numerous theoretical analyses have been performed to determine
precise global stellar parameters and test the various complicate
physical effects on the stellar structure and evolutionary theory
(Th\'{e}venin et al. 2002; Eggenberger et al. 2004b, 2005; Kervella
et al. 2004; Miglio \& Montalb\'{a}n 2005; Provost et al. 2004,
2006; Tang et al. 2008a, 2008b).

$\tau$ Ceti (HR 509, HD 10700) is a G8 V metal-poor star, belonging
to population II. Extensive analyse of this star have been performed
by many scientists who have provided different non-seismic
observational results (such as effective temperature $T_{eff}$ and
luminosity $L$), depending on the different methods used, i.e.
interferometry and spectroscopy. Teixeira et al. (2009) detected
solar-like oscillations on $\tau$ Ceti, identified some possible
existing frequencies, and obtained the large separation around
$\Delta \nu$ = 169 $\mu$Hz with HARPS. These seismological data will
provide a constraint on the fundamental parameters of $\tau$ Ceti.
Moreover, $\tau$ Ceti will be one of the most promising southern
asteroseismic targets of the seismology programme of Stellar
Observations Network Group (Metcalfe et al. 2010).

In this work, using a mixture of conventional and asteroseismic
observed constraints, we try to determine modelling parameters of
$\tau$ Ceti with YREC. The observational constraints available to
$\tau$ Ceti are summarized in Sect. 2, while the details of the
evolutionary models are presented in Sect. 3. The seismic analyses
are carried out in Sect. 4. Finally, the discussion and conclusions
are given in Sect. 5.


\section{Observational constraints}

\subsection{Non-asteroseismic observational constraints }

\begin{table}
\caption{Non-asteroseismic observational data of $\tau$ Ceti. }
\begin{tabular}{c c c}
\hline\hline
Observable & Value & Source \\


 \hline
 Effective temperature $T_{eff}$(K) & $5264\pm100$  & (1)   \\

 & $5525 \pm 12$ & (3) \\

\hline
Luminosity $L/L_{\odot}$&$0.52\pm0.03$& (4)\\

&$0.50\pm0.006$ & (3)\\

&$0.488\pm0.010$& (2)\\
\hline

Metallicity $[Fe/H]_{s}$&$-0.5\pm0.03$& (1)\\
\hline
Surface heavy-element\\
abundance $[Z/X]_{S}$&  $0.0073\pm 0.0005$& (5)\\
\hline

Radius $R/R_{\odot}$ & $0.773\pm0.024$ & (5)\\

\hline \hline


 \end{tabular}\\
  References.---(1) Soubiran et al. (1998) , (2) Teixeira et al. (2009), (3) Pijpers et al. (2003a), (4) Pijpers (2003b),  (5) this paper.
  \end{table}

The metallicity derived from observations is [Fe/H] =$-0.5\pm0.03$
(Soubiran et al. 1998). The mass fraction of heavy-elements, Z, was
derived assuming $\log [Z/X]\approx [Fe/H]+\log[Z/X]_{\odot}$, and
$[Z/X]_{\odot}=0.0230 $ (Grevesse and Sauval, 1998), for the solar
mixture. We can therefore deduce that $[Z/X]_{s}=0.0068-0.0078$.
 The radius, as an important parameter for constraining stellar
models, was first measured by Pijpers et al. (2003a) using
interferometry. They determined the radius of $\tau$ Ceti
corresponding to $0.773\pm 0.004_{(int.)}\pm
0.02_{(ext.)}R_{\odot}$. The measurement of the radius was then
improved by Di Folco et al. (2004) and Di Folco et al. (2007).
Finally, Di Folco et al. (2007) determined the radius
$R=0.790\pm0.005 R_{\odot}$. In our work, we use a large value of
radius $R$ = $0.773\pm0.024 R_{\odot}$ which includes all the
surrounding observational radius.

The effective temperature and luminosity of $\tau$ Ceti are both
derived from spectroscopy ($5264\pm100$K and
$0.52\pm0.03L_{\odot}$), and by ensuring that we reproduce the
measured radius ( $5525\pm12$ K, $0.500\pm0.006L_{\odot}$), using
interferometry ( Soubiran et al. 1998; Pijpers et al. 2003a, 2003b).
In addition the luminosity of a star can be obtained by combining
our knowledge of the magnitude and distance. For $\tau$ Ceti, the
apparent magnitude $V=3.50\pm0.01$, with the revised
 parallax, gives an absolute magnitude $M_{V}=5.69\pm0.01$. Teixeira et al. (2009) derived a luminosity for
  $\tau$ Ceti of $L/L_{\odot}$ = $0.488\pm0.010$, using bolometric
correction for $\tau$ Ceti $B.C.$ = $-0.17\pm0.02$ (Casagrande et
al. 2006) and adopting an absolute bolometric magnitude for the Sun
of $M_{bol, \odot}=4.74$ (Bessel et al. 1998).

 Using above different effective temperatures and luminosities, we
can obtain three error boxes,  which error box A  ($5525 \pm 12$ K,
$0.50 \pm 0.006$ $L_{\odot}$) are denoted by crosses, error box B
($5264 \pm 100$ K, $0.52 \pm 0.03$ $L_{\odot}$) denoted by
triangles, and error box C ($5264 \pm 100$ K, $0.488\pm0.010$
$L_{\odot}$) denoted by diamonds, shown in Fig. 1(d), respectively.
 Meanwhile, we decided to increase all errors by a factor of 1.5, so
that our calibration of the star is only weakly constrained by these
values.

  All non-asteroseismic
observational constraints are listed in Table 1.

\subsection{Asteroseismic constraints}

Solar-like oscillations of the G8V star $\tau$ Ceti were detected by
Teixeira et al. (2009) with the HARPS spectrograph. Thirty-one
individual modes are identified (see Table 1 in Teixeira et al.
2009). The large frequency separation is about $\Delta\nu=169$ $\mu
Hz$.

\section{Stellar models}

\begin{table}
\caption{Input parameters for model tracks.}

\begin{tabular}{c c c c}
\hline\hline
         & Minimum&     Maximum &\\
Variable & Value &  Value & $\delta$ \\
 \hline

 Mass $M/M_{\odot}$ & 0.770 & 0.795 & 0.005   \\
\hline

 Mixing length $\alpha$ & 0.8 & 1.8 & 0.2\\
 \hline

 Initial heavy  element&&&\\
 abundance $Z_{i}$ &0.001&  0.008 & 0.0005\\
\hline

Initial hydrogen &&&\\
abundance $X_{i}$&$0.70$& 0.75 & 0.01\\

 \hline

 \end{tabular}\\
 Note.---The value $\delta$ defines the increment between minimum
 and maximum parameter values used to create the model array.

   \end{table}

\begin{figure}
\includegraphics[angle=0,width=9cm,height=8cm]{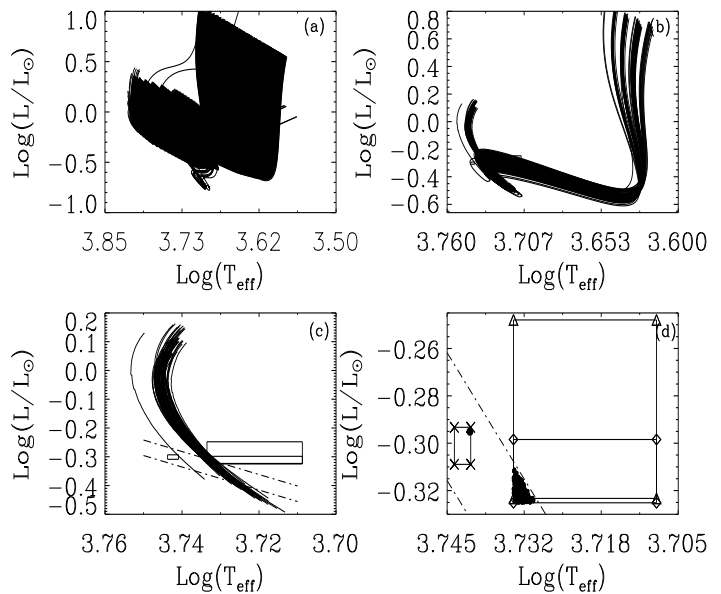}%

\caption{(a) All evolutionary tracks in the HR diagram; (b)
Evolutionary tracks falling in the error boxes from pre-main
sequence to main sequence; (c) Blow up the evolutionary tracks
falling in the error boxes in the main sequence; (d) The selected
models falling in the error boxes. Error box A  ($5525 \pm 12$ K,
$0.50 \pm 0.006$ $L_{\odot}$) is denoted by crosses, error box B
($5264 \pm 100$ K, $0.52 \pm 0.03$ $L_{\odot}$) denoted by
triangles, and error box C ($5264 \pm 100$ K, $0.488\pm0.010$
$L_{\odot}$) denoted by diamonds, respectively.}
\end{figure}

\begin{figure*}
\includegraphics[angle=0,width=9cm,height=8cm]{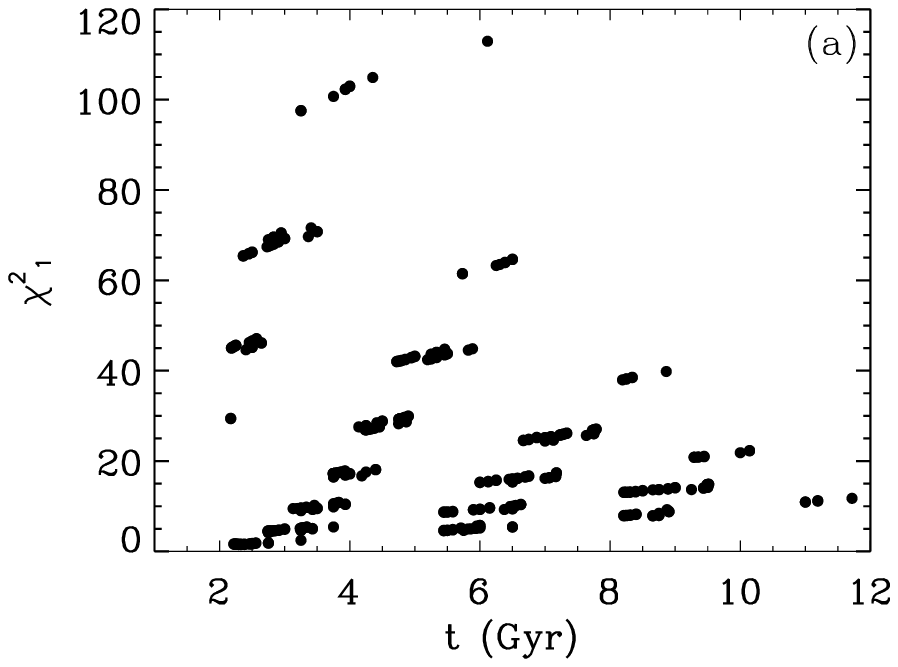}%
\includegraphics[angle=0,width=9cm,height=8cm]{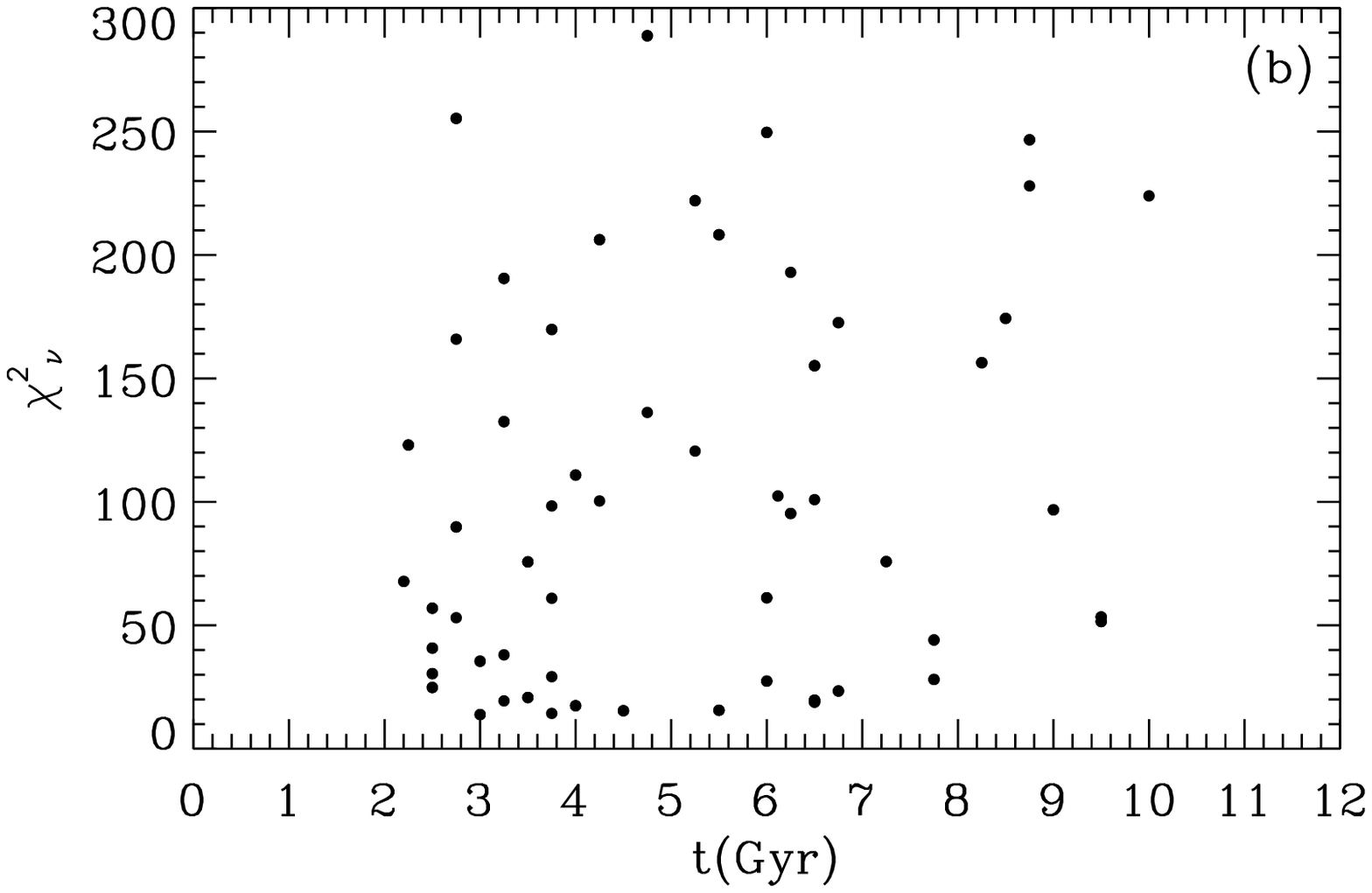}
\includegraphics[angle=0,width=9cm,height=8cm]{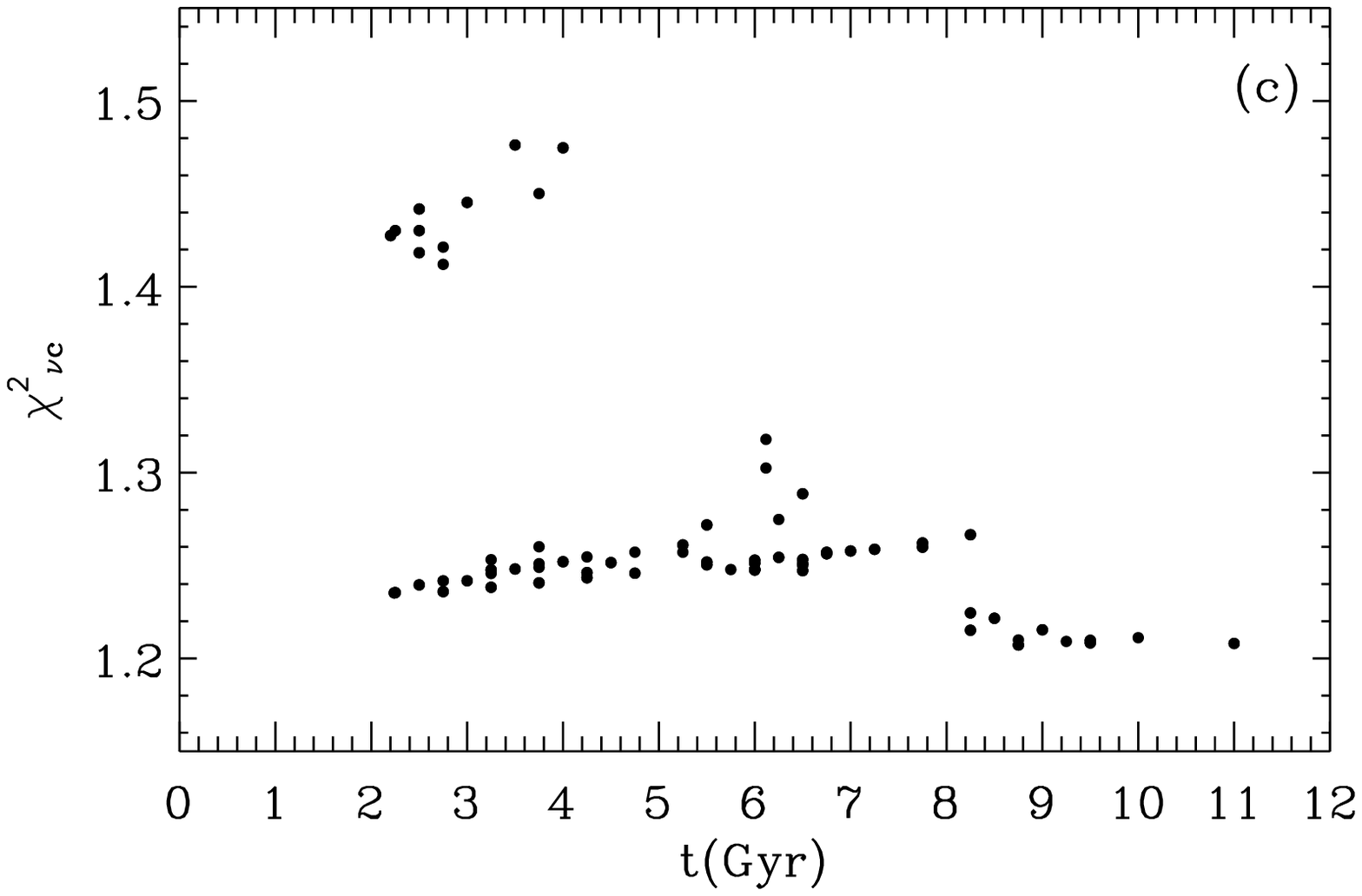}%
\includegraphics[angle=0,width=9cm,height=8cm]{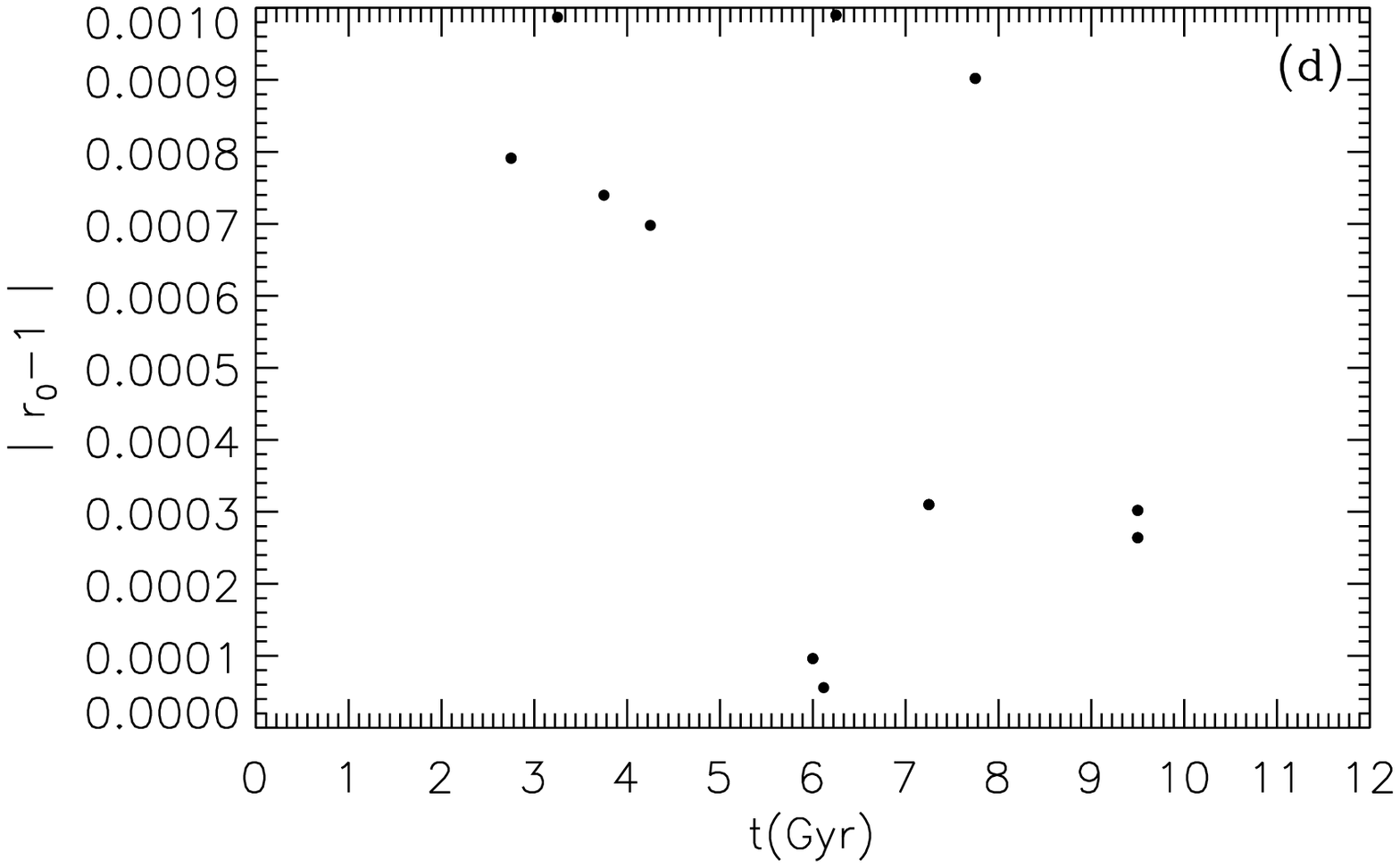}
\caption{(a) $\chi^{2}_{1}$ values derived from Eq.(2), plotted as a
function of age; (b) $\chi^{2}_{\nu}$ values derived from Eq.(3),
plotted as a function of age; (c) $\chi^{2}_{\nu c}$ values derived
from Eq.(6), plotted as a function of age; (d) $|r_{0}-1|$ values
 plotted as a function of age.}
\end{figure*}

 We calculated many evolutionary tracks using Yale stellar
evolution code (YREC; Demarque et al. 2008) by inputting different
parameters shown in Table 2.

The mass range are $M$ = 0.770 --
  0.795  $M_{\odot}$ with the increment value 0.005 $M_{\odot}$. Initial heavy element abundance range are $Z_{i}$ (0.001 -- 0.008) with
 the increment value 0.0005 and initial hydrogen abundance $X_{i}$ (0.70--0.75) with the
 increment value 0.01. Energy transfer by convection is treated according to the standard
mixing-length theory, and the boundaries of the convection zones are
determined by the Schwarzschild criterion (see Demarque et al. 2008
for details of the YREC). We set the mixing length parameter
$\alpha$=0.8--1.8 with the
 increment value 0.2. Using these parameter space, we created the model array. The
 initial zero-age main sequence (ZAMS) model
 used for $\tau$ Ceti is created from pre-main-sequence evolution calculations.
  These models are calculated using the updated OPAL equation-of-state
tables EOS2005 (Rogers and Nayfonov, 2002). We used OPAL high
temperature opacities (Iglesias and Rogers 1996) supplemented with
low temperature opacities from Ferguson et al. (2005). The NACRE
nuclear reaction rates (Angulo et al. 1999) were used.
 The Krishna-Swamy Atmosphere T-$\tau$ relation
is used for solar-like star (Guenther and Demarque 2000). All models
included gravitational settling of helium and heavy elements using
the formulation of Thoul et al. (1994).

\begin{figure}

\includegraphics[angle=0,width=9cm,height=8cm]{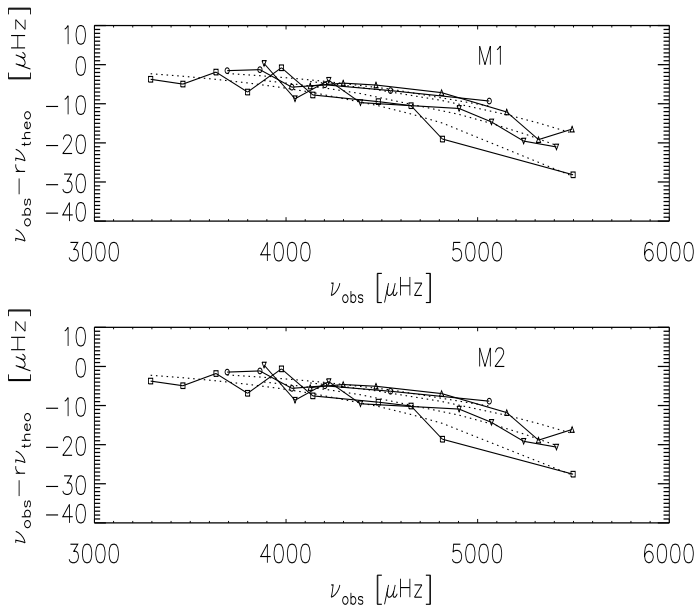}%
\caption{The difference between observed and best-fit model
frequencies, according to the left term of Eq.(4). Squares are used
for $l$ = 0 modes, diamonds for $l$ =1 modes, triangles for $l$ = 2
modes, and circles for $l$ = 3. Dotted lines show the power-law
function, according to the right term of Eq.(4).}
\end{figure}

Figure1(a) shows that many evolutionary tracks cover all possible
evolutionary status of $\tau$ Ceti. According to the above four
error boxes, we select all the tracks crossing the error boxes shown
in Fig.1(b).  We only choose to study main-sequence models , which
are shown in Fig.1(c). Meanwhile, we use the mass and radius to
estimate the large separation according to Eq. (1) ( Kjeldsen \&
Bedding 1995; Miglio et al. 2009). Furthermore, using the
temperature, luminosity, radius, and larger separation (refer to the
values from Teixeira et al. 2009) as constrainst, we select the
models of $\tau$ Ceti provided in Fig.1(d) as candidates.

\begin {equation}
\Delta \nu=\sqrt{\frac{M/M_{\odot}}{(R/R_{\odot})^{3}}}\times134.9  \mu Hz
\end {equation}

We now consider a function that describes the agreement between the
observations and the theoretical results
 \begin {equation}
\chi^{2}_{1}=\frac{1}{5}\sum^{5}_{i=1}(\frac{C^{theo}_{i}-C^{obs}_{i}}{\sigma
C^{obs}_{i}})^{2},
\end {equation}
where $\mathbf{C}$ represents the quantities $L/L_{\odot}$,
$T_{eff}$, $R/R_{\odot}$, and $[Fe/H]_{s}$ and large frequency
separation $\Delta \nu$, $\mathbf{C}^{theo}$ represents the
theoretical values, and $\mathbf{C}^{obs}$ represents the
observational values listed in Table 1. The vector $\sigma
\mathbf{C}_{i}^{obs}$ contain the errors in these observations,
which are also given in Table 1. We also decided to adopt a large
error (all errors are increased by a factor of 1.5), so that our
calibration of the star is only weakly constrained by these values,
which is not precisely determined. Figure 2(a) presents the values
$\chi^{2}_{1}$ versus age t of selected models that are shown in
Fig.1(d). We find that we cannot select an optimal model from
Fig.2(a). From Fig.2(a), we find that it is difficult to select an
optimal model depending mainly on the non-seismic constraints and
$\Delta\nu$, which was estimated by simply scaling from solar value
using Eq.(1). Hence, a detailed pulsation analysis are needed in the
next step.

\section{Asteroseismic constraints of fundamental parameters}

Using Guenther's pulsation code (Guenther 1994), we calculate the
adiabatic low-$l$ $p$-mode frequencies, the large- and  small-
frequency separations  ($\Delta
\nu_{n,l}\equiv\nu_{n,l}-\nu_{n-1,l}$ and $\delta
\nu_{n,l}\equiv\nu_{n,l}-\nu_{n-1,l+2}$, defined by Tassoul 1980) of
all the selected models. We compare the theoretical frequencies with
the corresponding observational frequencies using the function
$\chi^{2}_{\nu}$

\begin {equation}
\chi^{2}_{\nu}=
\frac{1}{N}\sum_{n,l}(\frac{\nu_{l}^{theo}(n)-\nu_{l}^{obs}(n)}{\sigma})^{2},
\end {equation}
 where, N=31 is the total number of modes, and $\nu_{l}^{theo}(n)$ and $\nu_{l}^{obs}(n)$ are the
 theoretical and observed frequencies respectively, for each spherical degree $l$ and the radial order $n$,
 where $\sigma = 2 \mu Hz$ (Teixeira et al. 2009) represents the uncertainty in the observed
 frequencies and $\chi^{2}_{\nu}$ values, plotted as function of age, are shown in Fig.2(b).

Since existing stellar models fail to accurately represent
 the near-surface layers of the solar-like stars, where the turbulent convection take place,
 the systematic offset between the observed and model frequencies appears. Furthermore,
 this offset between observed and best model frequencies turns out to
 be closely fitted by a power law (Christensen-Dalsgaard \& Gough 1980; Kjeldsen et al. 2008; Metcalfe et al. 2009; Do\v{g}an et al. 2009, 2010;
 Bedding et al. 2010; Christensen-Dalsgaard et al. 2010). In other words, this offset
  increases with increasing frequency shown in Fig.3. This power law can be expressed using the equation

 \begin {equation}
\nu_{obs}(n)-r_{l}\nu_{theo}(n)=a_{l}[\nu_{obs}(n_{i})/\nu_{max}]^{b},
\end {equation}
where $\nu_{obs}$ are the observed frequencies of radial and
non-radial order, $\nu_{best}=r_{l}\nu_{theo}(n)$ are the
corresponding calculated frequencies of the best-fit model, and
$\nu_{max}$ is a constant frequency corresponding to the peak power
in the spectrum,
 which is taken as 4490 $\mu Hz$ for $\tau$ Ceti and $r_{l}$, $a_{l}$, and $b$ a
 re parameters described in detail by Kjeldsen et al. (2008),
  (For a different spherical degree $l$,
   the values of $r$ and  $a$ are denoted by $r_{l}$ and $a_{l}$, respectively.).
    For the Sun and a solar-like star, the exponent $b=4.90$ is appropriate, as has been provn by many scientists.
     We use the Kjeldsen et al. (2008) prescription to correct the theoretical frequencies from near surface effects.

According to Eq. (4), we can use the following equation to obtain
the corrected frequencies of models:

 \begin {equation}
\nu_{correct}(n)=r_{l}\nu_{theo}(n)+a_{l}[\nu_{obs}(n)/\nu_{max}]^{b}.
\end {equation}

We define the function $\chi^{2}_{\nu c}$ in a similar way to Eq.(3)
as

\begin {equation}
\chi^{2}_{\nu c}= \frac{1}{N}\sum_{n,l}(\frac{\nu_{l}^{correct}(n)-\nu_{l}^{obs}(n)}{\sigma(\nu_{l}^{obs}(n))})^{2}.
\end {equation}
The values of  $\chi^{2}_{\nu c}$, plotted as a function of age are
shown in Fig.2(c). From Fig. 2(c), we can see that the values of
$\chi^{2}_{\nu c}$ are lower than $\chi^{2}_{\nu}$ and their lowest
values correspond to model ages from 8 to 10 Gyr. We conclude that
the optimal model corresponds to the lower values of
$\chi^{2}_{\nu}$ and $r_{0}-1$.
 From Figs.2(c) and 2(d), we infer that
only two models M1 and M2  can be accurately described by the
observational constraints. The difference between the observed and
uncorrected model frequencies of M1 and M2  are shown in Fig. 3. The
uncorrected and corrected frequencies of the optimal models M1 and
M2 and the observational frequencies are shown in Table 3.
\begin{figure*}
\includegraphics[angle=0,width=17cm,height=16cm]{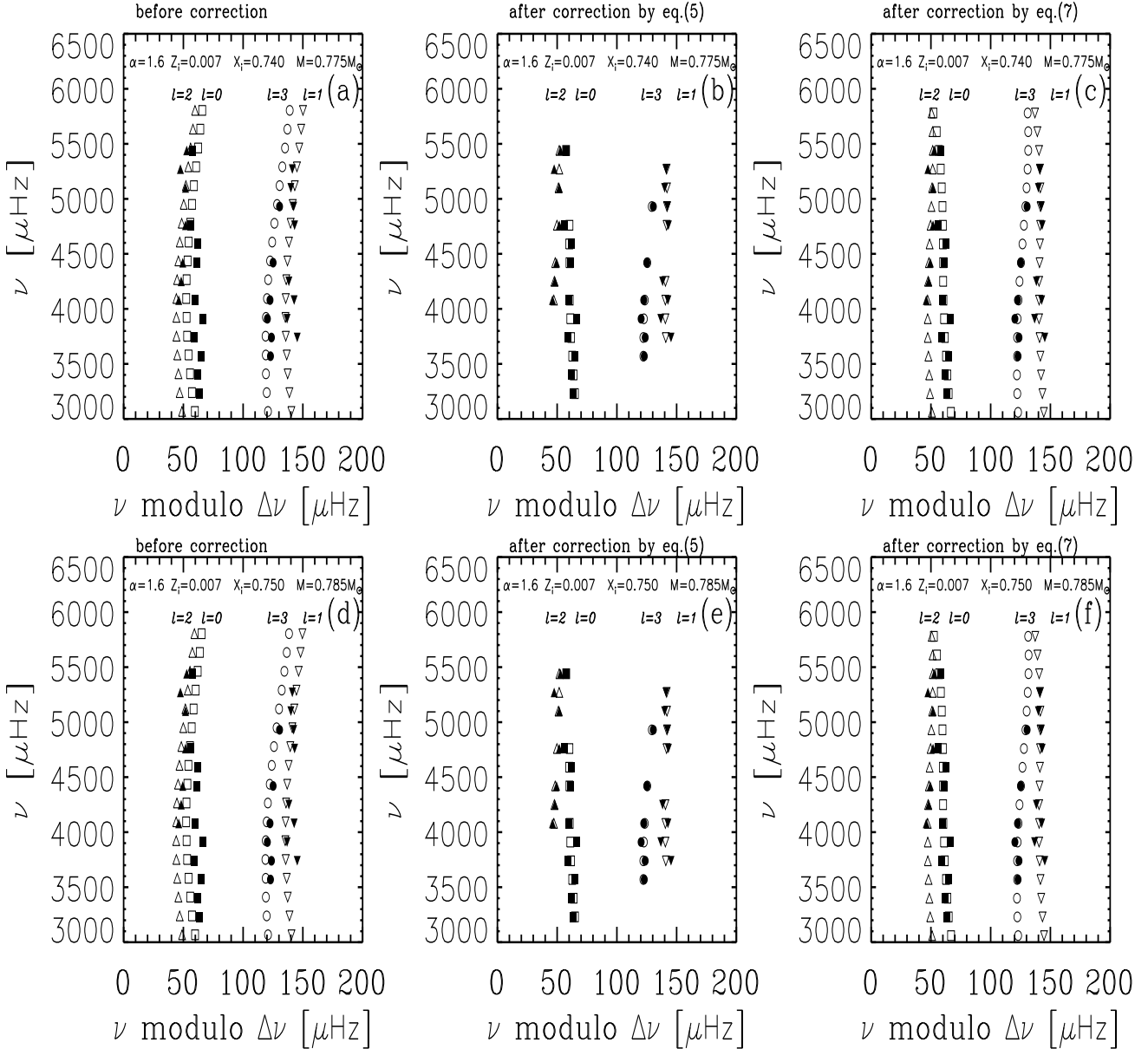}%
\caption{Echelle diagrams for the optimal models M1 (upper panel)
and M2 (lower panel). Left panel shows the case before applying
near-surface corrections. Middle panel
  shows the case after applying near-surface corrections, according to  Eq.(5). Right panel shows the case after
   applying near-surface
   corrections, according to  Eq.(7). Open symbols refer to
the theoretical frequencies, and filled symbols refer to the
observable frequencies. Squares are used for $l$ = 0 modes, diamonds
for $l$ =1 modes, triangles for $l$ = 2 modes, and
 circles for $l$ = 3. The observable frequencies
correspond to the average large separation about 170 $\mu Hz$ (see
text for details).}
\end{figure*}

\begin{figure*}

\includegraphics[angle=0,width=9cm,height=8cm]{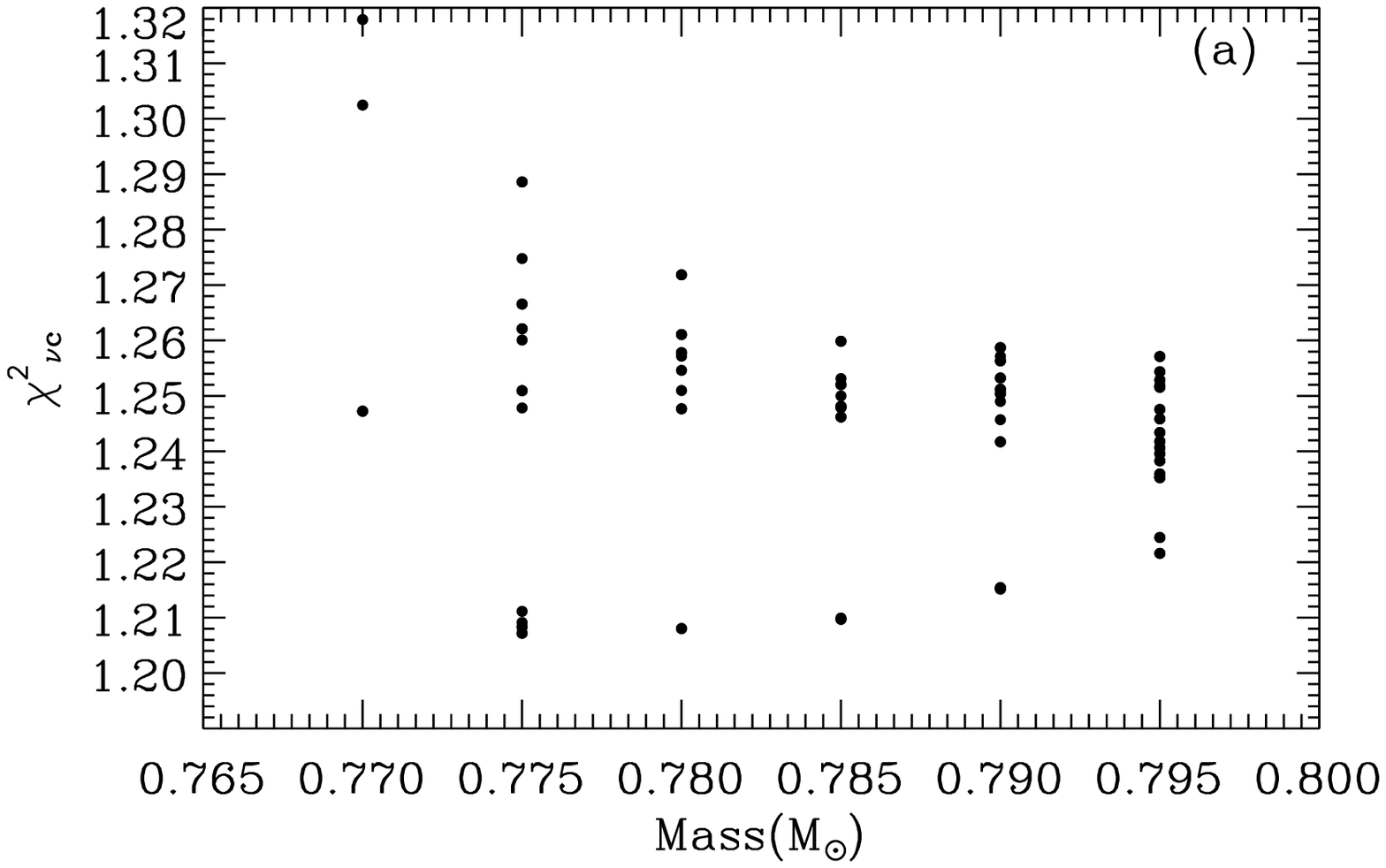}%
\includegraphics[angle=0,width=9cm,height=8cm]{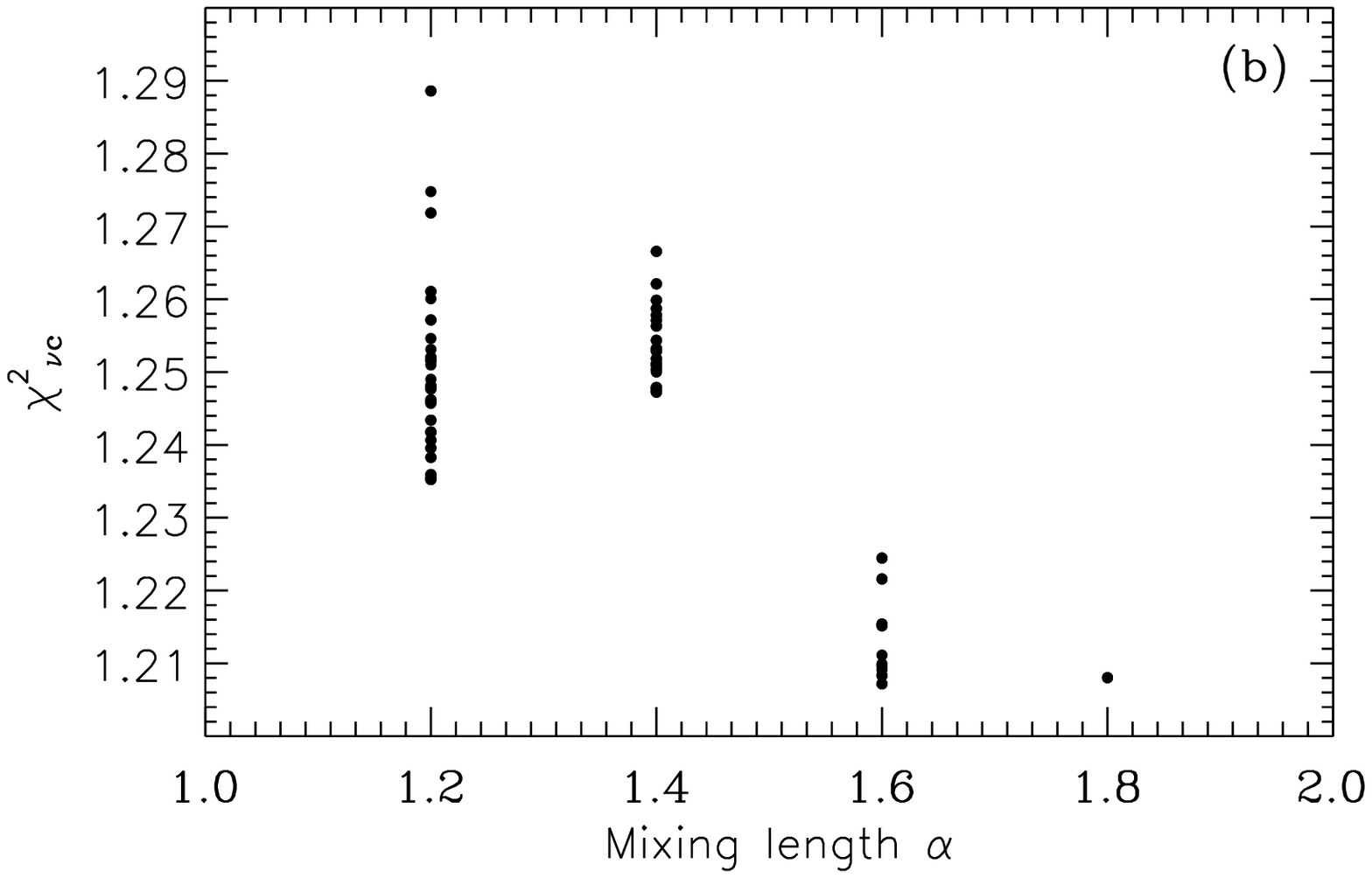}
\includegraphics[angle=0,width=9cm,height=8cm]{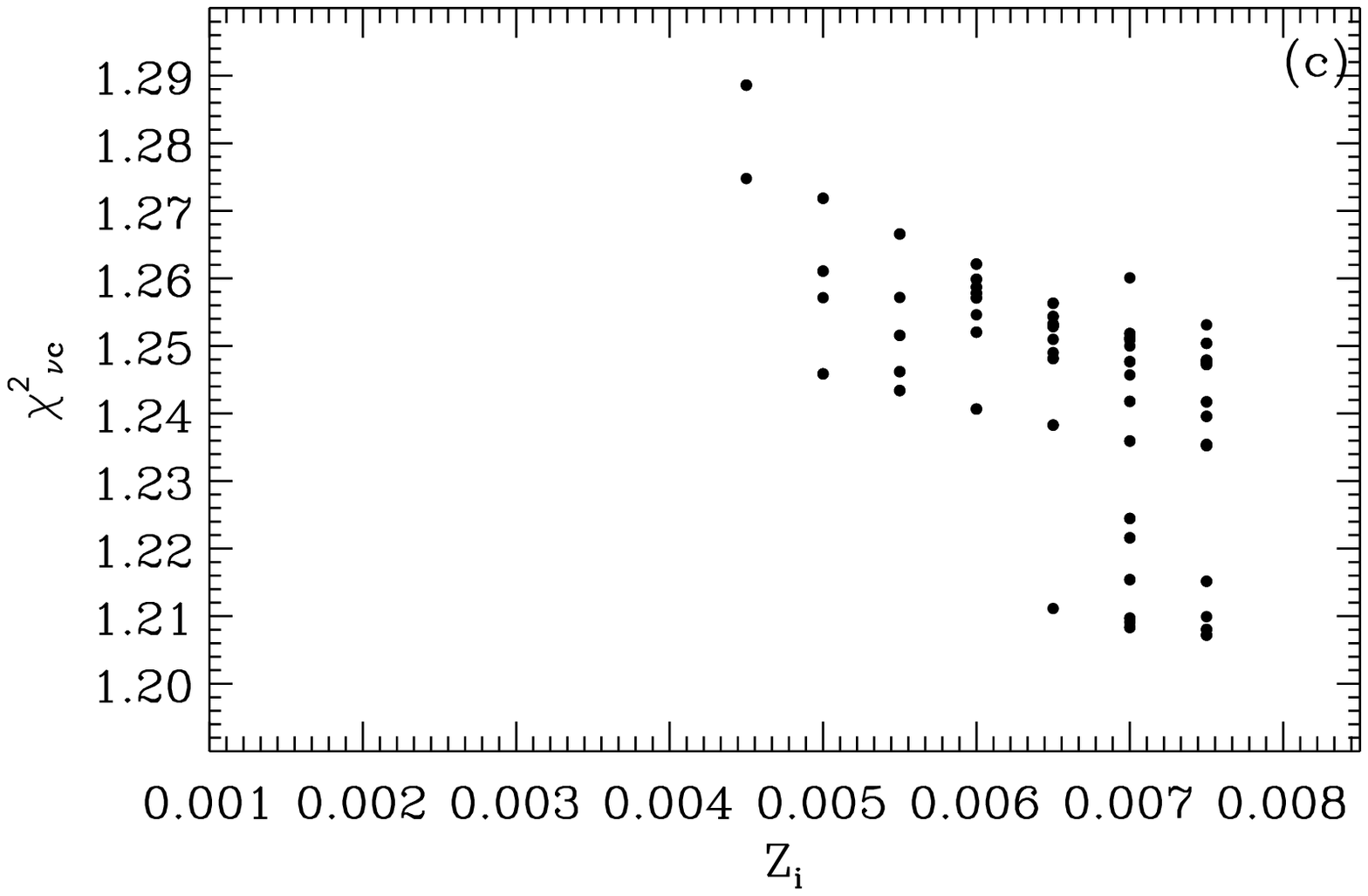}%
\includegraphics[angle=0,width=9cm,height=8cm]{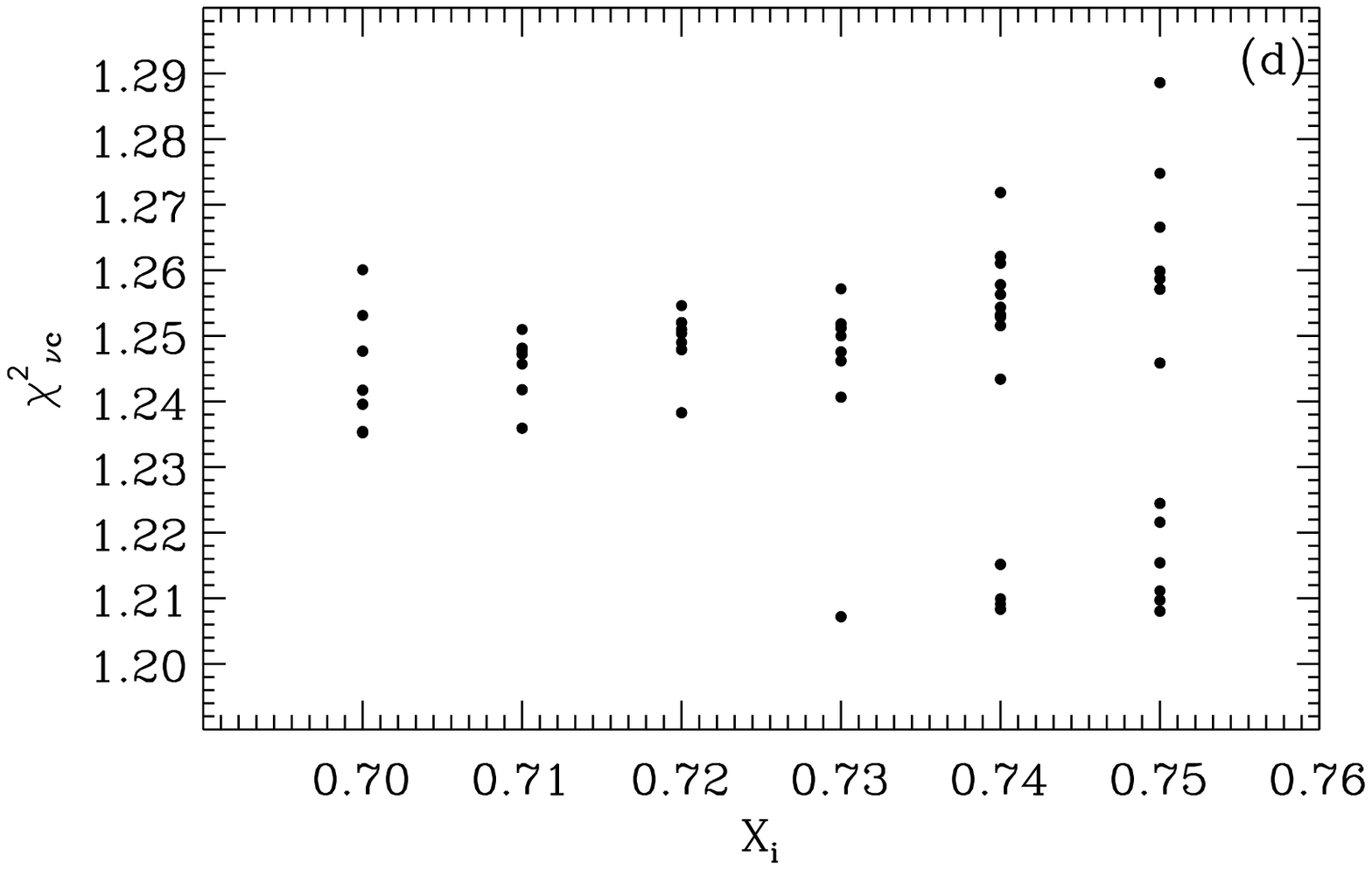}
\caption{(a). $\chi_{\nu c}^{2}$ values derived from Eq. (6),
plotted as function of mass; (b). $\chi_{\nu c}^{2}$ values plotted
as function of mixing length $\alpha$; (c). $\chi_{\nu c}^{2}$
values plotted as function of initial heavy element abundance
$Z_{i}$; (d). $\chi_{\nu c}^{2}$ values plotted as function of
initial hydrogen abundance $X_{i}$.}
\end{figure*}

To clearly compare all of the theoretical frequencies of the models
with observational frequencies, we provide echelle diagrams of
models M1 and M2 in Fig.4. An Echelle diagram is a useful tool
 for comparing stellar models with observations. This diagram presents the mode frequencies along the ordinate axis,
 and the same frequencies modulo the large separations in abscissae. From Figs.4(a) and 4(d), it can be seen that the
 uncorrected theoretical frequencies are not closely in agreement with the observed frequencies. The corrected theoretical
 frequencies indicated by  Eq. (5) fit perfectly the observation shown in Figs.4(b) and 4(e). Because the
 observed frequencies of orders $n$ are not consecutive and the values of $\nu_{obs}(n)$ are very close to those of
  $\nu_{theo}(n)$, we substitute the
  $\nu_{theo}(n)/\nu_{max}$ for $\nu_{obs}(n)/\nu_{max}$. Hence Eq.
  (5) becomes
 \begin {equation}
\nu_{correct}(n)=r\nu_{theo}(n)+a[\nu_{theo}(n)/\nu_{max}]^{b}.
\end {equation}
From Figs.4(b), 4(c), 4(e), and 4(f), it can be seen that corrected
frequencies given by  Eq. (5) and (7) respectively are uniform and
reproduce the observed frequencies perfectly. Furthermore, we can
use the function $\chi^{2}_{\nu c}$ to select the fitting model
parameters. As we all know, the suitable model parameters correspond
to the lowest values of $\chi^{2}_{\nu c}$, which can be clearly
seen in Fig. 5. From Fig. 5, we can conclude that the mass is in the
range $0.775 - 0.785 M_{\odot}$, $\alpha$ is in the range $1.6 -
1.8$, $Z_{i}$ in $0.0065 - 0.0075$, and $X_{i}$
 $0.73 - 0.75$. Hence, the model parameters of  $\tau$ Ceti can be constrained to within these narrow ranges.
Finally, we list the model parameters and characteristics of models
M1 and M2 in Table 4.

 \begin{table*}
\caption{The observational frequencies and the theoretical
frequencies for model M1 \& M2 before and after correction for
near-surface offset, respectively. }

 \tabcolsep0.05in
\tiny{ \[
\begin{tabular}{c c c c c c c c c c c c c}
\hline\hline
&&&&&before & &correction&&&&&\\
\hline
 &&  Observational & frequencies&&&  model & M1 &&  & model & M2& \\

 \hline

 $n$ & $l=0$ & $l=1$& $l=2$& $l=3$& $l=0$ & $l=1$& $l=2$& $l=3$ & $l=0$ & $l=1$& $l=2$& $l=3$    \\

 \hline
 18 &3293.4&...&...&...&3296.149&3377.700&3455.831&3529.092&  3296.276&3377.775&3455.826&3529.043  \\
\hline
 19 &3461.7&...&...&3692.9&3465.623&3547.268&3625.910&3699.994& 3465.717&3547.304&3625.854&3699.900  \\
 \hline
 20 &3634.5&...&...&3863.7&3635.309&3717.485&3796.205&3870.802& 3635.352&3717.479& 3796.119&3870.664 \\
  \hline
 21 &3799.3&3885.3&...&4030.3&3805.155&3887.715&3967.112&4042.136& 3805.169&3887.661&3966.987&4041.970\\
   \hline
 22 &3976.1&4046.8&4126.1&4202.5&3975.695&4058.363&4138.126&4213.984& 3975.674&4058.279&4137.957&4213.769\\
  \hline
 23 &4139.9&4222.7&4298.2&...&4146.398&4229.665&4309.760&4385.981&4146.331&4229.535&4309.557&4385.721\\
   \hline
 24 &...&4388.3&4469.5&4545.1&4317.694&4401.101&4481.820&4558.582&4317.594&4400.922&4481.566&4558.284\\
    \hline
 25&4481.8&...&...&...&4489.499&4573.112&4653.968&4731.322&4489.349&4572.896&4653.669&4730.972\\
     \hline
 26&4652.3&...&4811.8&...&4661.385&4745.381&4826.607&4904.208&4661.190&4745.115&4826.269& 4903.817\\
    \hline
 27&4816.2&4903.1&...&5060.5&4833.772&4917.748&4999.286&5077.435& 4833.537&4917.439&4998.898&5077.001\\
     \hline
 28&...&5072.3&5151.8&...& 5006.247&5090.515&5172.103&5250.549&5005.962&5090.165&5171.678&5250.071\\
   \hline
 29&...&5240.0&5317.5&...& 5178.835&5263.220&5345.147& 5423.822&5178.510&5262.825&5344.685& 5423.311\\
  \hline
30&...&5411.2&5492.8&...& 5351.685&5436.051& 5518.036&5597.086&5351.322&5435.623&5517.539&5596.541\\
  \hline
31&5497.9&...&...&...& 5524.391&5608.945&  5691.011&5770.097&5523.990&5608.485&5690.491&5769.528\\
\hline \hline
&&&&&after& &correction&&&&&\\
\hline
  18 &3293.4&...&...&...         &3294.811 &3373.293 &3448.558 & 3521.990 & 3294.873 & 3373.414 & 3448.676 & 3522.028\\
  \hline
   19 &3461.7&...&...&3692.9       &3463.687 &3542.188 &3617.891 & 3692.166 & 3463.725 & 3542.282 & 3617.973 & 3692.180\\
 \hline
 20 &3634.5&...&...&3863.7       &3632.637 &3711.616 &3787.344 & 3862.157 & 3632.638 & 3711.682 & 3787.412&  3862.151    \\
 \hline
 21 &3799.3&3885.3&...&4030.3    &3801.588 &3880.925 &3957.295 & 4032.565 & 3801.579 & 3880.961 & 3957.344 & 4032.561   \\
 \hline
 22 &3976.1&4046.8&4126.1&4202.5 &3971.050 &4050.500 &4127.228 & 4203.364 & 3971.028 & 4050.527 & 4127.255 & 4203.345   \\
 \hline
 23 &4139.9&4222.7&4298.2&...    &4140.467 &4220.554 &4297.635 & 4374.174 & 4140.428 & 4220.560 & 4297.652 & 4374.152   \\
 \hline
 24 &...&4388.3&4469.5&4545.1    &4310.239 &4390.548 &4468.303 & 4545.430 & 4310.201 & 4390.534 & 4468.298 & 4545.417   \\
 \hline
 25&4481.8&...&...&...           &4480.250 &4560.894 &4638.878 & 4716.651 & 4480.204 & 4560.877 & 4638.861 & 4716.641   \\
 \hline
 26&4652.3&...&4811.8&...        &4650.045 &4731.252 &4809.736 & 4887.824 & 4650.003 & 4731.224 & 4809.718 & 4887.837   \\
 \hline
 27&4816.2&4903.1&...&5060.5     &4820.001 &4901.433 &4980.409 & 5059.120 & 4819.977 & 4901.408 & 4980.383 & 5059.162   \\
 \hline
 28&...&5072.3&5151.8&...        &4989.674 &5071.706 &5150.968 & 5230.066 & 4989.668 & 5071.692 & 5150.953 & 5230.148   \\
 \hline
 29&...&5240.0&5317.5&...        &5159.046 &5241.582 &5321.475 & 5400.908 & 5159.079 & 5241.584 & 5321.477 & 5401.051   \\
\hline
30&...&5411.2&5492.8&...         &5328.222 &5411.212 &5491.524 & 5571.452 & 5328.308 & 5411.249 & 5491.552 & 5571.669   \\
\hline
31&5497.9&...&...&...            &5496.757 &5580.498 &5661.325 & 5741.433 & 5496.908 & 5580.579 & 5661.397 & 5741.746   \\
\hline \hline


 \end{tabular}\]}

  \end{table*}

\begin{table}
\caption{Final model-fitting results for $\tau$ Ceti. }
\begin{tabular}{c c c}
\hline\hline
Modelling parameters & model M1 & model M2 \\

 \hline
 Mass $M/M_{\odot}$ & 0.775  & 0.785   \\

 \hline
 Mixing length $\alpha$ &1.6  & 1.6   \\

\hline
$Z_{i}$&0.007&0.007\\

\hline

$X_{i}$&0.740& 0.750\\
\hline\hline

Model characteristics &  & \\
\hline

 Effective temperature $T_{eff}$(K) & 5409  & 5387   \\

\hline
Luminosity $L/L_{\odot}$&0.47985& 0.47612\\

\hline

Log(g)&4.53187& 4.53365\\
\hline

Radius $R/R_{\odot}$ & 0.78994 & 0.79339\\
\hline
$(Z/X)_{s}$ &0.00753&0.00749\\
\hline
Age (Gyr)& 9.5&9.5\\
\hline
$<\Delta \nu_{0}>$ ($\mu Hz$)&170.9222&170.9106\\
\hline
$<\Delta \nu_{1}>$ ($\mu Hz$)&170.8621&170.8381\\
\hline
$<\Delta \nu_{2}>$ ($\mu Hz$)&171.0555&171.0332\\
\hline
$<\Delta \nu_{3}>$ ($\mu Hz$)& 171.5120&171.4870\\
\hline
$<\delta \nu_{02}>$ ($\mu Hz$)&10.013&10.111\\
\hline
$<\delta \nu_{13}>$ ($\mu Hz$)&18.034&18.136\\
\hline \hline

 Model corrected paraments&&\\
 \hline
 $r_{0}$&1.000302&1.000264\\
 \hline
 $r_{1}$&0.9993002&0.9993007\\
 \hline
 $r_{2}$&0.9984142&0.9984387\\
 \hline
 $r_{3}$&0.9984967&0.9984996\\
 \hline
 $a_{0}$&-10.59438&-10.32439\\
 \hline
 $a_{1}$&-8.270579&-8.092409\\
 \hline
 $a_{2}$&-6.517972&-6.377440\\
 \hline
 $a_{3}$&-5.891401&-5.639216\\
 \hline \hline


 \end{tabular}\\
  \end{table}

\section{Discussion and conclusions}

Using the asteroseismic analysis and the empirical frequency
correction for the near-surface offset presented by Kjeldsen et al.
(2008) to correct our theoretical frequencies, we have derived the
optimal model of $\tau$ Ceti and now list our main conclusions:

1. Using the latest asteroseismic observations, we have attempted to
construct the optimal model of $\tau$ Ceti. We have only considered
the models M1 and M2 , which can closely describe the observations,
as the optimal models. Furthermore, the model parameters of $\tau$
Ceti have been constrained to within narrow intervals by the
function $\chi^{2}_{\nu c}$, where the mass is in the range $M$ =
$0.775 - 0.785 M_{\odot}$, the mixing length parameter in the range
$\alpha$ = $1.6 - 1.8$, the initial metallicity in the range $Z_{i}$
= $0.0065 - 0.0075$, the initial hydrogen abundance in the range
$X_{i}$ = $0.73 - 0.75$, and the age in the range $t$ = $8 - 10$
$Gyr$.

2. We have found that the results of the non-asteroseismic
observations (effective temperature and luminosity) inferred from
spectroscopy  are more accurate than those derived from
interferometry for $\tau$ Ceti, because our optimal models are in
the error boxes B and C derived from our spectroscopy results.

\vspace{2mm}

\begin{acknowledgements} 
We are grateful to the anonymous referee for his/her constructive
suggestions and valuable remarks that helped us to improve the
manuscript. We also thank Professor Shaolan Bi and Dr Linghuai Li
for many useful comments and discussions. This work was supported by
the support of Shandong Nature Science Foundation (ZR2009AM021),
Dezhou University Foundation(402811), and supported by The Ministry
of Science and Technology of the People¡¯s Republic of China through
grant 2007CB815406, and by NSFC grants 10773003, 10933002, and
10978010.
\end{acknowledgements}


\begin{thebibliography}{}

\bibitem[2008]{} Arentoft, T., Kjeldsen, H., Bedding, T. R., et al. 2008, ApJ, 687, 1080

\bibitem[1999]{} Angulo, C., Arnould, M., Rayet, M., et al. 1999,
Nucl. Phys. A., 656, 3

\bibitem[1998]{} Bessell, M. S., Castelli, F. \& Plez, B. 1998, A\&A, 333, 231

\bibitem[2001]{} Bedding, T. R., Butler, R. P., Kjeldsen, H., et al. 2001, ApJ, 549, L105

\bibitem[2004]{} Bedding, T. R., Kjeldsen, H., Butler, R.P., et al. 2004, ApJ, 614, 380
 \bibitem[2006]{} Bedding, T. R., Butler, R. P., Carrier, F. , et al. 2006, ApJ, 647, 558

\bibitem[2007]{} Bedding, T. R., Kjeldsen, H., Arentoft, T., et al. 2007, ApJ, 663, 1315
\bibitem[2010]{} Bedding, T. R., Kjeldsen, H., Campante, T. L., et al. 2010, ApJ, 713, 935

\bibitem[2002]{} Bouchy, F. \& Carrier, F. 2002, A\&A, 390, 205

\bibitem[2005]{} Bouchy, F., Bazot, M., Santos, N.C., Vauclair, S. \& Sosnowska, D. 2005, A\&A, 440, 609

\bibitem[2010]{} Bruntt, H., Bedding, T. R., Quirion, P. -O., et al. 2010, MNRAS, accepted


\bibitem[1991]{} Brown, T. M., Gilliland, R. L., Noyes, R. W., Ramsey, L. W. 1991, A\&A, 368, 599

\bibitem[2001]{} Carrier, F., Bouchy, F., Kienzle, F., et al. 2001, A\&A 378, 142

\bibitem[2003]{} Carrier, F. \& Bourban, G. 2003a, A\&A, 406, L23

\bibitem[2003]{} Carrier, F., Bouchy, F. \& Eggenberger, P., 2003b. In: Thompson, M.J., Cunha, M.S.,
Monteiro, M.J.P.F.G. (Eds.), Asteroseismology Across the HR Diagram. Kluwer, p. P311.


\bibitem[2005]{} Carrier, F., Eggenberger, P., DAlessandro, A. \& Weber, L., 2005a, NewA, 10, 315
\bibitem[2005]{} Carrier, F., Eggenberger, P. \& Bouchy, F. 2005b, A\&A 434, 1085

\bibitem[2006]{} Carrier, F.\& Eggenberger, P. 2006, A\&A, 450, 695

 \bibitem[2007]{} Carrier, F., Kjeldsen, H., Bedding, T. R., et al.  2007, A\&A, 470, 1059


 \bibitem[2010]{}Carrier, F., Morel, T., Miglio, A., et al. 2010, Ap\&SS, 328, 83
\bibitem[2006]{} Casagrande, L., Portinari, L.\& Flynn, C. 2006, MNRAS, 373, 13

\bibitem[1980]{}Christensen-Dalsgaard, J. \& Gough, D. O. 1980, Nature, 288, 544

 \bibitem[2010]{}Christensen-Dalsgaard, J., Kjeldsen, H., Brown, T. M., et al. 2010, ApJ, 713, L164


\bibitem[2010]{} Chaplin, W. J., Appourchaux, T., Elsworth, Y., et al. 2010, ApJ, 713, L169

 \bibitem[2006]{} De Ridder, J., Barban, C., Carrier, F., et al. 2006, A\&A, 448, 689

 \bibitem[2004]{} Di Folco, E., Th¨¦venin, F., Kervella, P., et al. 2004, A\&A, 426, 601

 \bibitem[2007]{} Di Folco, E., Absil, O., Augereau, J.-C., et al. 2007, A\&A, 475, 243

\bibitem[2009]{} Do\v{g}an, G., Brand\~{a}o, I. M., Bedding, T. R., et al. 2009, Ap\&SS. tmp..251D

\bibitem[2010]{} Do\v{g}an, G., Bonanno, A.\& Christensen-Dalsgaard, J. 2010, appear in the HELAS IV International Conference proceedings in Astronomische Nachrichten

\bibitem[2008]{} Demarque, P., Guenther, D. B., Li, L. H., et al. 2008, Ap\&SS, 316, 31

\bibitem[2004]{} Eggenberger, P., Carrier, F., Bouchy, F. \& Blecha, A. 2004a, A\&A 422, 247

\bibitem[2004]{} Eggenberger, P., Charbonnel, C., Talon, S., et al. 2004b, A\&A, 417, 235

\bibitem[2005]{} Eggenberger, P., Carrier, F. \& Bouchy, F. 2005, NewA, 10, 195

\bibitem[2008]{} Eggenberger, P., Miglio, A., Carrier, F., et al. 2008, A\&A, 482, 631

\bibitem[2005]{} Ferguson, J. W., Alexander, D. R., Allard, F., et al. 2005, ApJ, 623, 585

\bibitem[2008]{} Gai, N., Bi, S. L. \& Tang, Y. K. 2008, ChJAA, 8, 591

\bibitem[1994]{} Gray, D. F. \& Baliunas, S. L. 1994,  ApJ , 427, 1042-1047
 \bibitem[1998]{} Grevesse, N. \& Sauval, A. J. 1998,  SSRv, 85, 161

\bibitem[1992]{} Guenther, D. B., Demarque, P., Kim, Y.-C., et al. 1992, ApJ, 387, 372

\bibitem[1994] {} Guenther, D.B. 1994, ApJ 422, 400.

\bibitem[2000] {}Guenther, D.B.\& Demarque, P. 2000, ApJ 531, 503.

\bibitem[1996]{} Iglesias, C. A., \& Rogers, F. J. 1996, ApJ, 464, 943

\bibitem[2004]{} Judge, P. G., Saar, S. H., Carlsson, M., Ayres, T. R. 2004, ApJ, 609, 392

\bibitem[2010]{} Kallinger, T., Weiss, W. W., Barban, C., et al. 2010, A\&A, 509, id.A77

\bibitem[2004]{} Kervella, P., Th\'{e}venin, F., Morel, P., et al. 2004, A\&A, 413, 251

 \bibitem[1995]{} Kjeldsen, H. \& Bedding, T. R. 1995, A\&A, 293, 87

\bibitem[2003]{} Kjeldsen, H., Bedding, T.R., Baldry, I.K., et al. 2003, AJ, 126, 1483

\bibitem[2005]{} Kjeldsen, H., Bedding, T.R., Butler, R.P., et al. 2005, ApJ 635, 1281

\bibitem[2008]{} Kjeldsen, H., Bedding, T. R.\& Christensen-Dalsgaard, J. 2008, ApJ, 683, L175

\bibitem[2002]{} Li, L.H., Robinson, F.J., Demarque, P., Sofia, S. \& Guenther, D.B. 2002. ApJ, 567, 1192

\bibitem[2005]{}  Miglio, A. \& Montalb\'{a}n, J. 2005, A\&A, 441, 615

\bibitem[2009]{} Miglio, A.; Montalb¨¢n, J.. Eggenberger, P., et al. 2009, AIP Conference Proceedings, Volume 1170, pp. 132-136

\bibitem[2009] {} Miglio, A., Montalb¨¢n, J., Baudin, F., et al. 2009,  A\&A, 503, L21

\bibitem[2004] {}  Marti\'{c}, M., Lebrun, J.-C., Appourchaux, T. \& Korzennik, S.G. 2004a, A\&A, 418, 295

\bibitem[2004] {}Marti\'{c}, M., Lebrun, J.C., Appourchaux, T., Schmitt, J. 2004b, In: Danesy, D. (Ed.),
SOHO 14/GONG 2004 Workshop, Helio- and Asteroseismology: Towards a Golden Future, ESA SP-559. p. 563.

\bibitem[2009] {} Metcalfe, T. S., Creevey, O. L.\& Christensen-Dalsgaard, J. 2009, ApJ, 699, 373

\bibitem[2010]{} Metcalfe, T. S., Judge, P. G., Basu, S., et al. 2010, American Astronomical Society, AAS Meeting 215, 424.16

\bibitem[2005]{} Mosser, B., Bouchy, F., Catala, C., et al. 2005, A\&A, 431, L13

\bibitem[2003]{} Pijpers, F. P., Teixeira, T. C., Garcia, P. J., et al. 2003a, A\&A, 406, L15

\bibitem[2003]{} Pijpers, F. P. 2003b, A\&A, 400, 241

\bibitem[2004] {} Provost, J., Martic, M. \& Berthomieu, G. 2004, ESA SP 559, 594.

\bibitem[2006] {} Provost, J., Berthomieu, G., Marti\'{c}, M. \& Morel, P. 2006, A\&A 460, 759

\bibitem[2002]{} Rogers, F. J., \& Nayfonov, A. 2002, ApJ, 576, 1064

\bibitem[2003]{} Robinson, F.J., Demarque, P., Li, L. H., et al. 2003. MNRAS 340, 923.

\bibitem[1998]{} Soubiran, C., Katz, D. \& Cayrel, R. 1998, A\&AS, 133, 221

\bibitem[2007]{} Samadi, R., Georgobiani, D., Trampedach, R., et al. 2007, A\&A, 463, 297

\bibitem[2010] {} Stello, D., Chaplin, W. J., Basu, S., et al. 2010, MNRAS, 400, L80


\bibitem[2008]{} Tang, Y. K., Bi, S. L., Gai, N., et al. 2008a, ChJAA, 8, 421

\bibitem[2008] {} Tang, Y. K., Bi, S. L. \& Gai, N. 2008b, New Astronomy, 13, 541

\bibitem[1980]{} Tassoul, M. 1980, ApJS, 43, 469

\bibitem[2009] {} Teixeira, T. C., Kjeldsen, H., Bedding, T. R., et al. 2009, A\&A, 494, 237

\bibitem[2002] {} Th\'{e}venin, F., Provost, J., Morel, P., et al. 2002, A\&A, 392, 9

 \bibitem[1994]{} Thoul, A. A., Bahcall, J. N. \& Loeb, A. 1994. ApJ 421, 828.

%

\end{thebibliography}
\end{document}